\documentclass[letterpaper,twocolumn,amsmath,amssymb]{revtex4}
\usepackage{siunitx}
\usepackage{nidanfloat}
\usepackage{graphicx}
\usepackage{here}
\usepackage{booktabs}
\usepackage{setspace}
\usepackage[svgnames]{xcolor}
\usepackage[unicode,colorlinks=true,citecolor=blue,linkcolor=blue,urlcolor=NavyBlue]{hyperref}
\setcitestyle{numbers,comma,sort&compress}
\usepackage{url}
\usepackage[normalem]{ulem}
\usepackage[nameinlink,capitalize]{cleveref}
\usepackage{mhchem}
\begin{document}
\title{Flux trapping in hydrides not yet confirmed}
\author{N. Zen}
\affiliation{National Institute of Advanced Industrial Science and Technology,\\Tsukuba Central 2-10, Ibaraki 305-8568, Japan}
\begin{abstract}
In Ref.~\cite{natphys}, Minkov et al reported the time dependence of magnetic moment of hydride materials under high pressure in a diamond anvil cell. Here we point out that the straight lines interpolated with the measurement results give the misleading impression that thermal flux creep in superconductors decays linearly with time. To dispel this misconception, we correctly interpolated logarithmic decay curves. Then, it has become apparent that there is no reason to assume that the measured magnetic moment originates from flux trapping, i.e., there is no reason to believe that persistent currents are circulating in hydride materials under high pressure and that they are superconductors.{\if0 Instead, we argue that the assumption originates from confirmation bias.\fi}
\end{abstract}
\maketitle
To prove that newly created materials are superconductive, flux trapping experiments are as crucial as demonstrating the Meissner effect, as well summarized in Ref.~\cite{Sridhar2023}.{\if0 Both are physical phenomena related to the behaviour of magnetic flux within a material.\fi} The latter is a phenomenon where magnetic flux is expelled from the interior of a material becoming superconducting, while the former is a phenomenon where trapped magnetic flux within a superconducting material remains persistent indefinitely over cosmic timescales. Therefore, the time dependence of thermally activated flux creep in superconductors is always logarithmic. Because it is logarithmic, trapped flux does not die out over cosmic timescales. This is well explained in an introductory textbook for superconductivity~\cite{TinkhamBook} and has been well verified experimentally, e.g., in Refs.~\cite{Kim1962,Esquin2012,Eley2017,Thompson2005}.

\begin{center}
\begin{figure*}[h!b]
  \hspace{-20mm}
  \includegraphics[width=0.73\textwidth]{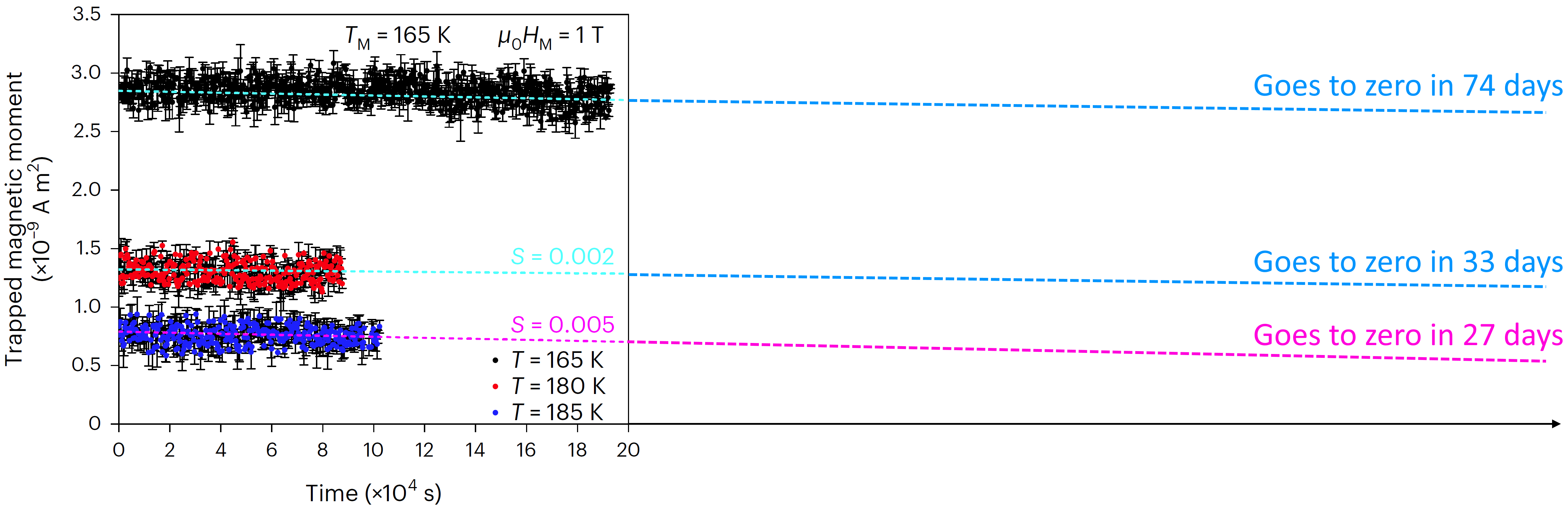}
  \caption{Fig.~4c of Ref.~\cite{natphys} showing time-decay of magnetic moment. Minkov et al interpolated straight lines in the graph, and we have extended them.}
  \label{fig1}
  \vspace{2ex}
  \centering
  \includegraphics[width=0.83\textwidth]{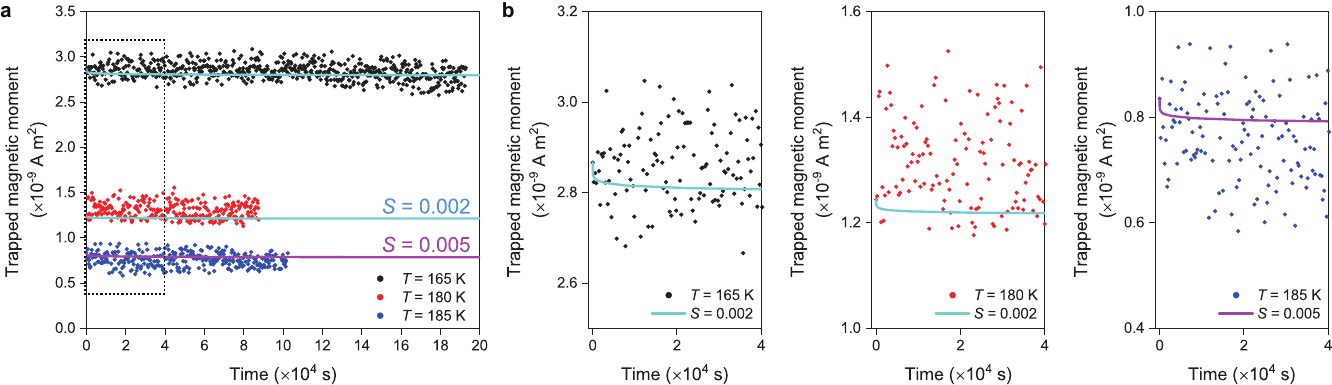}
  \caption{\textbf{a}, Raw data for Fig.~4c of Ref.~\cite{natphys} are plotted. For readability, error bars displayed in Fig.~4c of Ref.~\cite{natphys} are omitted. In the graph, logarithmic decay curves instead of linear ones are correctly interpolated. The fundamental feature of the logarithmic decay function, namely, decreasing rapidly at first and later becoming more and more gradual, can be confirmed by zooming-in the region surrounded by the dotted rectangle, as shown in \textbf{b}: the left, middle, and right panels are the data for 165 K, 180 K, and 185 K, respectively. In none of these plotted data, can we infer, let alone claim, a logarithmic decay with time.}
  \label{fig2}
\end{figure*}
\end{center}

Note that in Ref.~\cite{natphys} Minkov et al cite Refs.~\cite{Eley2017,Thompson2005}. Therefore, they should be aware that thermal flux creep decays logarithmically with time. Nevertheless, they interpolated straight lines in the measurement data, as shown in \cref{fig1}, taken from Fig.~4c of Ref.~\cite{natphys}. If magnetic moment decays linearly with time as shown, it will go to zero in less than a year in any case. Since flux trapping by superconductivity cannot die out in such a short time, these interpolated straight lines must be corrected with logarithmic curves, as shown in \cref{fig2}.
{}
\onecolumngrid
\twocolumngrid
The flux creep rate $S$ is given by
\begin{equation}
  S=-\frac{dln(m_{trap})}{dln(t)}\approx -\frac{1}{m_{trap,0}} \frac{dm_{trap}}{dln(t)},\label{eq:S}
\end{equation}
where $m_{trap}$ is the magnetic moment at elapsed time $t$ and $m_{trap,0}$ is that at $t=0$. (In Ref.~\cite{natphys}, the minus sign in this formula is accidentally missing.) Using the publicly available magnetic moment raw data at $t=0~\si{s}$ for $m_{trap,0}$ and the $S$ values indicated in Fig.~4c of Ref.~\cite{natphys}, we can obtain formulae for $m_{trap}$ as a function of $t$, which is of course a logarithmic decay function, and we have interpolated them in the magnetic moment-versus-time graph, as shown in \cref{fig2}a. The plots in the graph were made from the publicly available raw data and are therefore essentially the same as Fig.~4c of Ref.~\cite{natphys}; only error bars are omitted for readability. Although the correctly interpolated logarithmic curves still look like straight lines, the fundamental feature of the logarithmic decay function, namely, decreasing rapidly at first and later becoming more and more gradual, can be confirmed in the zoomed-in graphs shown in \cref{fig2}b.

Then, it has become apparent that there is no hint of logarithmic decay feature anywhere in the measured magnetic moment. If one were to discover logarithmic trajectories within such featureless scatter plots, it would require an exceptionally strong belief which is unrelated to science. This is a typical example of how strong beliefs can overshadow experimental results. Feynman left a quote, `\emph{The first principle is not to fool yourself ...}'. Yet his voice has not reached the exceptionally strong belief that lies at the core of hydride materials, the BCS theory.

\begin{figure}[H]
\centering
\includegraphics[width=0.95\linewidth]{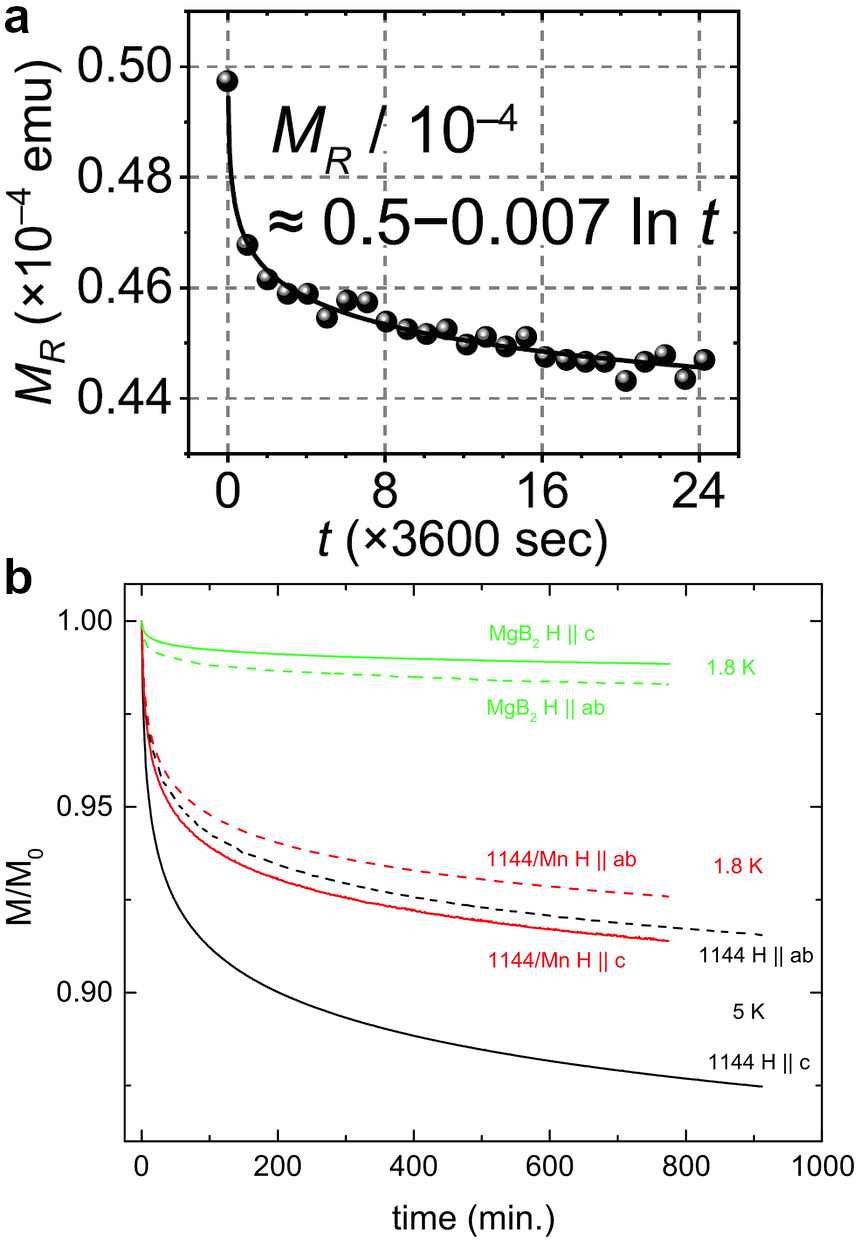}
\caption{Time dependence of remnant magnetisation of superconducting samples: \textbf{a}, taken from Fig.~6 of Ref.~\cite{sayonarabcs} reported by the author of this paper; \textbf{b}, taken from Fig.~A9 of Ref.~\cite{Budkosust} reported by one of the authors of Ref.~\cite{natphys}. In all the cases shown here, logarithmic decay can be observed by anyone. In Fig.~4c of Ref.~\cite{natphys}, by contrast, no one would be able to observe a logarithmic decay without confirmation bias.}
\label{fig3}
\end{figure}

\Cref{fig3} shows typical examples of the time dependence of remnant magnetisation of superconducting samples measured in the absence of an applied magnetic field. \Cref{fig3}a is taken from Fig.~6 of Ref.~\cite{sayonarabcs} reported by the author of this paper, and \cref{fig3}b is taken from Fig.~A9 of Ref.~\cite{Budkosust} reported by one of the authors of Ref.~\cite{natphys}. As can be easily confirmed, the time dependence of remnant magnetisation is \emph{always} logarithmic, no matter which material is measured, no matter who measures it, as long as the material is truly superconducting.{\if0 Someone may fool themselves, but no one can fool this universal feature.\fi}
{}

In contrast, the universal feature of logarithmic decay is totally absent in Fig.~4c of Ref.~\cite{natphys}. One of the authors of Ref.~\cite{natphys} states in the abstract of Ref.~\cite{Budkosust}, `\emph{Our results, on these known superconductors at ambient pressure, are qualitatively similar to those recently measured on superhydrides at megabar pressures (Minkov {\it et al} 2023 {\it Nat. Phys.} https://doi.org/10.1038/s41567-023-02089-1) and, as such, hopefully serve as a baseline for the interpretation of high pressure, trapped flux measurements.}' However, there is no `qualitative similarity' between Fig.~A9 of Ref.~\cite{Budkosust} and Fig.~4c of Ref.~\cite{natphys}, i.e., between \cref{fig3}b and \cref{fig1} of this paper. The former exhibits the fundamental feature of logarithmic decay, namely, decreasing rapidly at first and later becoming more and more gradual, while the latter lacks it entirely. If they truly believe that there is a qualitative similarity, it would have been appropriate to interpolate logarithmic decay curves in Fig.~4c of Ref.~\cite{natphys} from the beginning, rather than linear ones. Even if they had done so, however, it is unlikely that their hope `\emph{to clear the doubts in the interpretation of the measurements in \cite{natphys}}' stated in Ref.~\cite{Budkosust} would become reality. On the contrary, we have found the traces of confirmation bias deeply etched in \cref{fig2}b. There is nothing wrong with making an assumption in advance. However, if experiments indicate there isn't, then simply there isn't. If the assumption has fallen apart, let it be. If it remains unbroken, it should indeed be exceptionally strong. No one is at fault. It is a frightening power of BCS theory.

In conclusion, we argue that the magnetic moments measured in Ref.~\cite{natphys} cannot be interpreted as magnetic moments of trapped flux, since there is no hint of logarithmic decay feature anywhere in the time dependence data in Fig.~4c of Ref.~\cite{natphys}. If one considers that the trapped magnetic moment is just buried within a large background, it is necessary to understand the magnetic behaviour of the background first. A possibility of the background being ferromagnetic has been suggested by Hirsch et al~\cite{HMtrapping}. If that is the case, it is necessary to obtain the background data first and then perform a proper background subtraction. Note that that should be done with great care because hydride materials under high pressure easily induce confirmation bias, as proven in this paper. In any case, whether the background is ferromagnetic or not, the vertical axes for the magnetic moment data presented in Ref.~\cite{natphys} cannot be labelled as `Trapped magnetic moment' at present. Namely, Figs.~1c, 1d, 4a, 4b, 4c and 5b of Ref.~\cite{natphys} lack accuracy and are therefore currently invalid. Figures~2 and 3 of Ref.~\cite{natphys}, which explicitly indicate the use of `trapped' flux in their captions, are of course invalid too.

In Ref.~\cite{natphys}, Minkov et al state, `\emph{We reveal that the absence of a pronounced Meissner effect is associated with the very strong pinning of vortices inside the samples.}' By subtracting the fact that flux trapping in hydride materials has not yet been confirmed from their statement, only the fact of the `\emph{absence of a pronounced Meissner effect}' remains. That is to say, the data shown in Figs.~1a and 5a of Ref.~\cite{natphys} cannot be assumed to be a pronounced Meissner signal. Thus, all the hydride materials under high pressure reported in Ref.~\cite{natphys}, namely \ce{H3S} and \ce{LaH10}, show neither a pronounced Meissner effect nor flux trapping, i.e., they do not meet the established criteria for a conclusive demonstration of superconductivity at all, which are defined in Ref.~\cite{Sridhar2023}.
\\
\\
\noindent
{\small
{\bf Data availability}\\
Raw data for \cref{fig3}a is available at \cite{Zenodo2022}.
\\
{\bf Competing interests}\\
None declared.
}
\urlstyle{same}


\begin{thebibliography}{99}
\bibitem{natphys}
  V. S. Minkov et al, ``Magnetic flux trapping in hydrogen-rich high-temperature superconductors'', \href{https://doi.org/10.1038/s41567-023-02089-1}{{\it Nat. Phys.} {\bf 19} 1293--1300 (2023)}.
\bibitem{Sridhar2023}
  S. Sridhar, ``True superconductivity at near ambient temperature has not been confirmed by Dasenbrock-Gammon et al. Nature, volume 615, pages 244--250 (2023)'', \href{https://doi.org/10.1016/j.jpcs.2023.111381}{{\it J. Phys. Chem. Solids} {\bf 180} 111381 (2023)}.
\bibitem{TinkhamBook}
  M. Tinkham, {\it Introduction to Superconductivity 2nd ed.}, Dover Publications, New York (1996).
\bibitem{Kim1962}
  Y. B. Kim et al, ``Critical Persistent Currents in Hard Superconductors'', \href{https://doi.org/10.1103/PhysRevLett.9.306}{{\it Phys. Rev. Lett.} {\bf 9} 306--309 (1962)}.
\bibitem{Esquin2012}
  T. Scheike et al, ``Can Doping Graphite Trigger Room Temperature Superconductivity? Evidence for Granular High-Temperature Superconductivity in Water-Treated Graphite Powder'', \href{https://doi.org/10.1002/adma.201202219}{{\it Adv. Mater.} {\bf 24} 5826--5831 (2012)}.
\bibitem{Eley2017}
  S. Eley et al, ``Universal lower limit on vortex creep in superconductors'', \href{https://doi.org/10.1038/nmat4840}{{\it Nat. Mater.} {\bf 16} 409--413 (2017)}.
\bibitem{Thompson2005}
  J. R. Thompson et al, ``Vortex pinning and slow creep in high-$J_{c}$ \ce{MgB2} thin films: a magnetic and transport study'', \href{https://doi.org/10.1088/0953-2048/18/7/008}{{\it Supercond. Sci. Technol.} {\bf 18} 970--976 (2005)}.
\bibitem{sayonarabcs}
  N. Zen, ``Sayonara BCS: Realization of Room Temperature Superconductivity as a result of a First Order Phase Transition'', \href{https://doi.org/10.48550/arXiv.2306.13172}{arXiv:2306.13172v3 (Submission: 14 September 2023)}.
\bibitem{Budkosust}
  S. L. Bud'ko et al, ``Trapped flux in pure and Mn-substituted \ce{CaKFe4As4} and \ce{MgB2} superconducting single crystals'', \href{https://doi.org/10.1088/1361-6668/acf413}{{\it Supercond. Sci. Technol.} {\bf 36} 115001 (2023)}.
\bibitem{HMtrapping}
  J. E. Hirsch et al, ``On the interpretation of flux trapping experiments in hydrides'', \href{https://doi.org/10.48550/arXiv.2309.14952}{arXiv:2309.14952v1 (Submission: 26 September 2023)}.
\bibitem{Zenodo2022}
  See ``Figure6\_MH-Mt24hours-MH.csv'' at {\it Zenodo} \href{https://doi.org/10.5281/zenodo.5885550}{https://doi.org/10.5281/zenodo.5885550} (Publication: 15 July 2022), uploaded by N. Zen.
\end{thebibliography}
\end{document}